\documentstyle[preprint,aps]{revtex}
\begin{document}
\preprint{}
\title
{Density nonlinearities and a field theory \\
for the dynamics of simple fluids
}
\author{ Gene F. Mazenko and Joonhyun Yeo }
\address{
The James Franck Institute and the Department of Physics,\\
The University of Chicago, Chicago, Illinois 60637
}
\maketitle
\begin{abstract}
We study the role of the Jacobian arising from a
constraint enforcing the nonlinear relation: ${\bf g}=\rho{\bf V}$,
where $\rho,\: {\bf g}$ and ${\bf V}$ are the mass density,
the momentum density and the local velocity field, respectively,
in the field theoretic formulation of the nonlinear
fluctuating hydrodynamics of simple fluids. By investigating the
Jacobian directly and by developing a field theoretic formulation
without the constraint, we find that no changes in dynamics result
as compared to the previous formulation developed by Das and Mazenko (DM).
In particular, the cutoff mechanism discovered by DM
is shown to be a consequence of the $1/\rho$ nonlinearity
in the problem not of the constraint. The consequences of this result
for the static properties of the system is also discussed.
\end{abstract}

\section{Introduction}
\label{sec:intro}
The development of the appropriate field theoretical
treatment
for nonlinear fluctuating hydrodynamics (NFH)
of simple fluids is a more
subtle enterprise than one might first imagine. If
one develops a theory for simple fluids including
the complete set of conserved fields,
the mass density $\rho({\bf x})$,
the momentum density
${\bf g(x)}$, and
the energy density $\varepsilon({\bf x})$,
then one
finds \cite{KM}
that one must include multiplicative noise \cite{Morozov}
in order to gain consistency with thermodynamics.
This surprising result is associated with the connection
between the fluctuating energy, entropy and
temperature. In the case where the energy is not
included in the set of slow variables, one still finds
technical problems in developing
the associated field theory. These
problems are associated with the form of the kinetic
energy density ${\bf g}^{2}/2\rho$
resulting from Galilean invariance. This $1/\rho$
factor
can be identified with the nonlinear relationship
between the momentum density and the local velocity
field: ${\bf V(x)}={\bf g(x)}/\rho({\bf x})$
and, at first sight, seems to complicate
the problem considerably.

In Ref.~\cite{DM} (DM), the
NFH of compressible fluids was studied
as a model for the glass transition. A field
theoretic formulation of the problem was developed by generalizing
the standard Martin-Siggia-Rose (MSR) method
\cite{MSR} to include the nonlinear
constraint between ${\bf g(x)}$ and ${\bf V(x)}$.
In the functional integral formulation described in
Ref.~\cite{DM}, this
constraint was enforced by introducing an auxiliary velocity field
${\bf V(x)}$ and inserting a delta-functional constraint,
\begin{equation}
\int\,{\cal D}{\bf V(x)}\:\delta ({\bf g}-\rho {\bf V}).\label{const}
\end{equation}
Enforcing the relation ${\bf g}=\rho {\bf V}$ eliminates the
$1/\rho $
nonlinearity in the kinetic energy
and yields a {\it polynomial} action in
$\rho ({\bf x}), {\bf g(x)}$ and ${\bf V(x)}$, which in turn
allowed one to carry out a perturbation theory expansion with
standard renormalization schemes. The
effects due to nonlinearities on
various physical quantities were then calculated
at the one-loop order and, as a
result of a systematic perturbative expansion, it was
discovered
that there is a nonhydrodynamic correction that cuts off the
sharp nature of the idealized glass transition
\cite{leut,GT}.
This result has important consequences that dense fluid
systems remain
ergodic for all values of the density and temperature,
although the density feedback mechnism
does drive the
viscosity to large values.

Recently Schmitz, Dufty and De \cite{duf} suggested that
a more faithful representation of the original Langevin
equations requires that the constraint should be
enforced by using,
instead of Eq.~(\ref{const}),
the identity
\begin{eqnarray}
1&=&\int\,\cal{D}{\bf V(x)}\:\delta(\frac{{\bf g}}{\rho}-{\bf V})
\label{const2} \\
&=&\|\rho\|\int\,\cal{D}{\bf V(x)}\:\delta ({\bf g}-\rho {\bf V}),
\label{jacob}
\end{eqnarray}
where $\|\rho\|$ is the Jacobian resulting from the change of
variable in the delta-functional. It was argued
in Ref.~\cite{duf} that
the cutoff mechanism found by DM may be an
artifact resulting from their use of the
constraint, Eq.~(\ref{const}) rather than
Eq.~(\ref{const2}). Since these constraints
differ only by the Jacobian factor
$\|\rho\|$, this is equivalent to saying
that the cutoff mechanism is eliminated
if one includes the Jacobian factor into the
development. We show here that this
is not the case.

We can evaluate the Jacobian $\|\rho\|$ directly
by expressing it in terms of an
integral over Grassmann fields \cite{grassmann}.
Inserting this result into the DM action, we
will be able to show that it does not play
any significant role on the dynamics.
We then develop a similiar functional integral
formulation to the one in Ref.~\cite{DM}, but without
introducing any
constraint. The $1/\rho$ nonlinearities are expanded as a
power series in $\delta\rho =\rho -\rho_{0}$, and
by comparing our
perturbation theory
results with those of Ref.~\cite{DM}, we will
be able to see explicitly the role of the Jacobian.
We find
that the two formulations are
equivalent at the
one-loop order of the perturbation theory
and the Jacobian has no dynamical
effects at this order.
This seems perplexing since it is clear that
the Jacobian does influence the static equilibrium
behavior generated by a general effective Hamiltonian.
The resolution to this apprent contradiction is
that the dynamics of a simple fluid, generated
by a Langevin equation, is insensitive to
changes in the chemical potential. It appears,
at least to lowest order, that the effects of the
Jacobian in question can be absorbed in the
chemical potential and does not directly
affect the statics generated by the Langevin equation.

In Sec.~\ref{rev}, we give a brief review of the formulation in
Ref.~\cite{DM}
just to collect some results needed for our discussion.
In Sec.~\ref{gras},
the Jacobian is expressed as a Grassmann integral
and the
usefulness of
this formulation is discussed. In Sec.~\ref{nov}, the theory
is formulated without
introducing the velocity field and the
one-loop equivalence between the two formulations are
studied in detail. The static theory is described in Sec.~\ref{sta}.

\section{Brief Review of DM}
\label{rev}

Our starting point is the set of
generalized Langevin equations for
compressible fluids for the set
of slow variables
$\{\rho({\bf x}), {\bf g(x)}\}$.
Following standard procedure \cite{MM}, we obtain
the continuity
equation for the conservation of mass
\begin{equation}
\frac{\partial\rho}{\partial t}=-\nabla\cdot{\bf g},
\end{equation}
and the generalized Navier-Stokes equation for
the conservation of momentum,
\begin{equation}
\frac{\partial g_{i}}{\partial t}=-\rho\nabla_{i}
\frac{\delta F_{u}}{\delta\rho}-\sum_{j}\nabla_{j}(\frac{
g_{i}g_{j}}{\rho})-\sum_{j}L_{ij}(\frac{g_{j}}{\rho})+\Theta_{i}.
\label{nav:stok}
\end{equation}
In Eq.~(\ref{nav:stok}),
$F_{u}[\delta\rho]$ is the potential energy part of the effective
Hamiltonian given by
\begin{equation}
F=F_{K}+F_{u},
\label{hamil}
\end{equation}
where the kinetic energy $F_{K}$ is
\begin{equation}
F_{K}=\int\,d^{d}{\bf x}\:\frac{{\bf g^{2}}}{2\rho}.
\label{kinetic}
\end{equation}
In Ref~\cite{DM}, the simple choice of $F_{u}$,
\begin{equation}
F_{u}[\delta\rho]=\int\,d^{d}{\bf x}\:\frac{A}{2}
(\delta\rho )^{2}
\label{potential}
\end{equation}
was used, where $A^{-1}\equiv\rho_{0}/c^{2}_{0}$
is the flat static
structure factor, $\rho_{0}$ is the
average mass density and $c_{0}$ is the bare
sound speed.
This simple quadratic part of $F_{u}$
then gives the
pressure nonlinearity
in the Langevin equation
which is responsible for the density
feedback mechanism of the glass transition \cite{leut}.
In general, $F_{u}$ can be any local functional
of $\delta\rho$ and the
spatial derivatives of $\delta\rho$.
By including the derivatives of $\delta\rho$,
we can probe the effect of
the spatial correlations in the system on the dynamics.
The dissipative matrix in Eq.~(\ref{nav:stok}) is given by
\begin{equation}
L_{ij}({\bf x})
=-\eta_{0}(\frac{1}{3}\nabla_{i}\nabla_{j}+\delta_{ij}
\nabla^{2})-\zeta_{0}
\nabla_{i}\nabla_{j},
\end{equation}
where $\eta_{0}$ is the bare shear viscosity and $\zeta_{0}$
the bare
bulk viscosity.
For later use we define the bare longitudinal viscosity
$\Gamma_0=\zeta_{0}+\frac{4}{3}\eta_{0}$.
The noise $\Theta_{i}$ is Gaussian with variance
\begin{equation}
\langle\Theta_{i}({\bf x},t)\Theta_{j}
({\bf x^{\prime}},t^{\prime})\rangle =
2k_{B}TL_{ij}({\bf x})\delta ({\bf x-x^{\prime}})
\delta (t-t^{\prime}).
\end{equation}

These Langevin equations can be put into a field theoretical form
following the standard MSR procedure \cite{MSR,MSR2}.
It essentially
amounts to introducing a hatted variable $\hat{\psi}$ for each
field
$\psi=\rho ,{\bf g}$ to enforce the equation of motion and
integrating over the Gaussian noise to yield a quadratic action
in $\hat{\psi}$. The generating functional without
source terms is given by
\begin{equation}
Z=\int\,{\cal D}\psi\,{\cal D}\hat{\psi}\:e^{-S[\psi,
\hat{\psi}]},
\label{Z1}
\end{equation}
where the action $S[\psi,\hat{\psi}]$, with the
notation $1=({\bf x_{1}},t_{1})$, is given
by
\begin{eqnarray}
S[\psi,\hat{\psi}]=\int &&\, d\,1\:\{\sum_{ij}\hat{g_{i}}
\beta^{-1}L_{ij}(1)
\hat{g_{j}}
+i\hat{\rho}[\frac{\partial\rho}{\partial t_{1}}+
\nabla_{1}\cdot {\bf g}] \nonumber \\
&&~~+i\sum_{i}\hat{g_{i}}[\frac{\partial g_{i}}{
\partial t_{1}}+\rho\nabla^{i}_{1}\frac{\delta F_{u}}
{\delta\rho}+\sum_{j}\nabla^{j}_{1}(\frac{g_{i}
g_{j}}{\rho})
+\sum_{j}L_{ij}(1)(\frac{g_{j}}{\rho})]\: \}. \label{a1}
\end{eqnarray}
In Ref.~\cite{DM}, the $1/\rho$ nonlinearities are removed by
using Eq.~(\ref{const}) and the representation of a
delta-functional,
\begin{equation}
\delta({\bf g}-\rho{\bf V})=\int\,{\cal D}\hat{{\bf V}}\:
\exp [i\int\,d1\:\hat{{\bf V}}(1)\cdot
({\bf g}(1)-\rho (1){\bf V}(1))],
\end{equation}
so that the generating functional in Ref.~\cite{DM} can
be expressed as the
functional integral over $\Psi $ and $\hat{\Psi}$, where
$\Psi =\rho, g_{i}, V_{i}$. Denoting hereafter
the results from Ref.~\cite{DM} by the supersript I, we have
\begin{equation}
Z^{(I)}=\int\,{\cal D}\Psi\,{\cal D}\hat{\Psi}\:e^{-S^{(I)}[\Psi,
\hat{\Psi}]},
\label{Z2}
\end{equation}
where $S^{(I)}[\Psi,\hat{\Psi}]$ is given by
\begin{eqnarray}
S^{(I)}[\Psi,\hat{\Psi}]=\int &&\, d\,1\:\{\sum_{ij}\hat{g_{i}}
\beta^{-1}L_{ij}(1)\hat{g_{j}}
+i\hat{\rho}[\frac{\partial\rho}{\partial t_{1}}+
\nabla_{1}\cdot {\bf g}] \nonumber \\
&&~~+i\sum_{i}\hat{g_{i}}[\frac{\partial g_{i}}{
\partial t_{1}}+\rho\nabla^{i}_{1}\frac{\delta F_{u}}
{\delta\rho}+\sum_{j}\nabla^{j}_{1}(\rho V_{i}V_{j})
+\sum_{j}L_{ij}(1)V_{j}] \nonumber \\
&&~~+i\hat{{\bf V}}\cdot({\bf g}-\rho{\bf V})\}. \label{a2}
\end{eqnarray}

The perturbative expansion for a polynomial action of this
form is standard.
The nonlinear corrections are given in the form of the
self-energies
$\Sigma_{\alpha\beta}$
modifying
the inverse linear propagator
$[G^{0}]^{-1}_{\alpha\beta}$
through Dyson's equation,
\begin{equation}
G^{-1}_{\alpha\beta}=[G^{0}]^{-1}_{\alpha\beta}
-\Sigma_{\alpha\beta},
\label{dyson}
\end{equation}
where $G_{\alpha\beta}=\langle\Psi_{\alpha}\Psi_{\beta}
\rangle $.
Detailed calculations of the self-energies
were carried out in Ref.~\cite{DM}.
Here we list the results for the density response
and the correlation functions.
In the Fourier-transformed space,
\begin{eqnarray}
&&G^{(I)}_{\rho\hat{\rho}}({\bf q},\omega )
=\frac{\rho({\bf q},\omega )\omega+iL({\bf q},\omega )}
{D^{(I)}({\bf q},\omega )},
\label{rrh1} \\
&&G^{(I)}_{\rho\rho}({\bf q},\omega )
=\frac{2\beta^{-1}q^{2}
\rho^{2}({\bf q},\omega )\tilde{L}({\bf q},\omega )}
{|D^{(I)}({\bf q},\omega )|^{2}},
\label{rr1}
\end{eqnarray}
where
\begin{eqnarray}
&&\rho({\bf q},\omega )=\rho_{0}-i\Sigma^{(I)}_{\hat{V}V}
({\bf q},\omega ), \label{r0}  \\
&&L({\bf q},\omega )=q^{2}\Gamma_{0}+i\Sigma^{(I)}_
{\hat{g}V}({\bf q},\omega ), \\
&&\tilde{L}({\bf q},\omega )=q^{2}\Gamma_{0}-(2\beta^{-1})
^{-1}\{ \Sigma^{(I)}_{\hat{g}\hat{g}}({\bf q},\omega )
+(\frac{\Gamma_{0}q^{2}}{\rho_{0}})^{2}
\Sigma^{(I)}_{\hat{V}\hat{V}}({\bf q},\omega )\},
\end{eqnarray}
and
\begin{equation}
D^{(I)}({\bf q},\omega )=\rho({\bf q},\omega )\{\omega
^{2}-q^{2}c^{2}_{0}-\Sigma^{(I)}_{\hat{g}\rho}
({\bf q},\omega )\}
+iL({\bf q},\omega )\{\omega+iq\Sigma
^{(I)}_{\hat{V}\rho}({\bf q},\omega )\}.
\label{D1}
\end{equation}
Here all the self-energies are their longitudinal parts.
The density feedback mechanism is realized by calculating
the one-loop self-energies contributing to the dynamic viscosities,
$L({\bf q},\omega )$ and $\tilde{L}({\bf q},\omega )$, which
are \cite{DM} quadratic in the density correlation
function \cite{leut}.
Without the nonhydrodynamic correction due to the self-energy
$\Sigma
^{(I)}_{\hat{V}\rho}({\bf q},\omega )$, the response
function is the same as the one that gives the density
feedback mechanism. Thus $\Sigma
^{(I)}_{\hat{V}\rho}({\bf q},\omega )$ provides the DM
cutoff. As mentioned in Sec.~\ref{sec:intro}
and as indicated by the subscripts, the cutoff
is related to the
$1/\rho$ nonlinearities and the
constraint between ${\bf g}$ and ${\bf V}$.

\section{Evaluation of the Jacobian}
\label{gras}

The question arises: what changes result if one
uses the constraint Eq.~(\ref{const2}) rather than Eq.~(\ref{const})
to introduce the velocity field? It should be clear from
Eq.~(\ref{jacob}) that the only changes in the
DM action comes from the Jacobian factor
$\|\rho\|$. This quantity
can be represented as
\begin{equation}
\|\rho\| =\int\,{\cal D}{\bf \eta}\,{\cal D}
{\bf \bar{\eta}}\:\exp [\int\,d1\:\sum_{i}\bar{\eta}_{i}
(1)\rho(1)\eta_{i}(1)],
\label{j:1}
\end{equation}
where $\bar{\eta}_{i}$ and $\eta_{i}$ are
Grassmann fields.
We must, as indicated below, be careful in using any unregularized
representation of the Jacobian like Eq.~(\ref{j:1}) in the functional
integral formalism, since it might not preserve causality.
If we include this in the DM formulation, then
the appropriate generating functional is given by
\begin{equation}
\tilde{Z}=\int\,{\cal D}{\bf \eta}\,{\cal D}
{\bf \bar{\eta}}\,{\cal D}\Psi\,{\cal D}\hat{\Psi}\:
e^{-\tilde{S}[\Psi, \hat{\Psi},{\bf \eta},{\bf \bar{\eta}}]},
\label{Z3}
\end{equation}
where the action is given now by
\begin{equation}
\tilde{S}[\Psi, \hat{\Psi},{\bf \eta},{\bf \bar{\eta}}]=S^{(I)}
[\Psi, \hat{\Psi}]+\int\,d1\:\sum_{i}\bar{\eta}_{i}
(1)\rho(1)\eta_{i}(1).
\label{a3}
\end{equation}
The only linear propagator involving ${\bf \bar{\eta}}$ or
${\bf \eta}$ is
\begin{equation}
G^{0}_{\bar{\eta}_{i}\eta_{j}}({\bf q},\omega )=
-G^{0}_{\eta_{j}\bar{\eta}_{i}}({\bf q},\omega )=
\frac{\delta_{ij}}{\rho_{0}}
\end{equation}
The new nonlinear term arising from the Grassmann fields is
of the form, $\delta\rho\sum_{i}\bar{\eta}_{i}\eta_{i}$. This
gives, at one-loop order, only two new self-energy
diagrams (Fig.~1). The first appears to give a
contribution to $\Sigma_{\rho\rho}$, which clearly violates
causality. We note that
the self-energies between two unhatted variables
$\Sigma_{\psi\psi}$ are indeed
equal to zero\cite{psipsi}
in the original DM formulation in accordance with the causality.
Thus Eq.~(\ref{j:1}) has to be regularized.
We note that if we use, instead of
Eq.~(\ref{j:1}),
\begin{equation}
\int\,d1\:\sum_{i}\bar{\eta}_{i}(1)\{\epsilon\frac
{\partial}{\partial t_{1}}+\rho (1)\}\eta_{i}(1),
\label{reg}
\end{equation}
with a time derivative with a very small coeffcient
$\epsilon$,
then
the first diagram of Fig.~1 vanishes,
since it is in the form of a time
integral of the product of a retarded and an advanced propagator.
Furthermore, as will be clear later in the section, the regularization
given as Eq.~(\ref{reg}) guarantees the causality to all orders
of perturbation theory, i.e. $\Sigma_{\psi\psi}$ remains to vanish
to all orders.

The second diagram in Fig.~1, however, does
give a finite contribution.
In the limit $\epsilon\rightarrow 0$, we have
\begin{equation}
\Sigma_{\bar{\eta_{i}}\eta_{j}}({\bf q},\omega )=
-\Sigma_{\eta_{j}\bar{\eta_{i}}}({\bf q},\omega )=
-\delta_{ij}\frac{\beta^{-1}}{c^{2}_{0}}\delta_{\Lambda}
(0)\equiv\delta_{ij}\Sigma,
\label{sigmaeta}
\end{equation}
where $\delta_{\Lambda}(0)=\int^{\Lambda}\,\frac{d^{d}{\bf k}}
{(2\pi )^{d}}$ with the large momentum
cutoff $\Lambda$. This self-energy
gives a correction to the correlation function
between $\eta$ and $\bar{\eta}$;
\begin{equation}
G_{\bar{\eta}_{i}\eta_{j}}({\bf q},\omega )=
\frac{\delta_{ij}}{\rho_{0}-\Sigma}\: .
\label{eta}
\end{equation}
We note that there are no
such self-energies at one-loop order
that link the Grassmann fields to
the $\rho ,{\bf g}$ or ${\bf V}$.
This fact, together with the vanishing of the first
diagram in Fig.~1, shows that the inverse propagator
matrix
in Eq.~(\ref{dyson}) is in a block-diagonal form
with the $\Psi\hat{\Psi}$ block being identical
to those given in DM formulation.
Inverting the inverse propagator,
we find that there are no changes, at one-loop order,
in the response
and correlation functions for $\rho, {\bf g}$ and
${\bf V}$ due to the Jacobian.

This result, in fact,
can be generalized to all orders of perturbation
theory by deriving the Ward identity from the
invariance of the action, Eq.~(\ref{a3}) under
$\eta_{i}\rightarrow e^{i\sigma}\eta_{i}$ and
$\bar{\eta}_{i}\rightarrow e^{-i\sigma}\bar{\eta}_{i}$
for any constant $\sigma$. Infinitesimally this symmetry
transformation reads
\begin{equation}
\delta\eta_{i}=(i\sigma )\eta_{i},~~~~~
\delta\bar{\eta}_{i}=(-i\sigma )\bar{\eta}_{i},~~~~~
\delta\Psi_{\alpha}=\delta\hat{\Psi}_{\alpha}=0.
\label{sym}
\end{equation}
Introducing the sources $J_{\alpha},\,\hat{J}_{\alpha}$
for $\Psi_{\alpha},\,\hat{\Psi}_{\alpha}$ and the
Grassmann sources $\bar{\xi}_{i},\,\xi_{i}$ for
$\eta_{i},\,\bar{\eta}_{i}$ respectively into the
generating functional, Eq.~(\ref{Z3}), we have
\begin{eqnarray}
\tilde{Z}[J,\hat{J},\bar{\xi},\xi ]&=&
\int\,{\cal D}{\bf \eta}\,{\cal D}
{\bf \bar{\eta}}\,{\cal D}\Psi\,{\cal D}\hat{\Psi}\:
\exp [\,
-\tilde{S}[\,\Psi, \hat{\Psi},{\bf \eta},{\bf \bar{\eta}}],
\nonumber \\
&& ~~~~~~+\int\, d1\: [\,\sum_{\alpha}\{ J_{\alpha}
\Psi_{\alpha}+\hat{J}_{\alpha}\hat{\Psi}_{\alpha}\}
+\sum_{i}\{\bar{\xi}_{i}\eta_{i}+\bar{\eta}_{i}\xi_{i}\}\, ]
\: ].
\label{sources}
\end{eqnarray}
Since the action $\tilde{S}$ and the integration measure
in Eq.~(\ref{sources}) are invariant under
the trasformation, Eq.~(\ref{sym}), the variation of the
source terms must vanish if we change
the integration variable by Eq.~(\ref{sym}). This
yields the following Ward identity:
\begin{equation}
\int\, d1\: \sum_{i}\,\{\xi_{i}(1)\frac{\delta}
{\delta\xi_{i}(1)}-\bar{\xi}_{i}(1)\frac
{\delta}{\delta\bar{\xi}_{i}(1)}\}\:
\tilde{Z}[J,\hat{J},\bar{\xi},\xi ]=0
\label{ward}
\end{equation}
Taking a derivative of Eq.~(\ref{ward}) with
respect to $\xi_{j}(3)$ and then
with respect to $J_{\alpha}(2)$ and setting all the
sources equal to zero, we have
\begin{equation}
0=\left.\frac{\delta^{2}}{\delta J_{\alpha}(2)\delta
\xi_{j}(3)}\,\tilde{Z}[J,\hat{J},\bar{\xi},\xi ]\right|
_{J=\hat{J}=\xi =\bar{\xi}=0}=-\tilde{Z}[0]\,G_{\Psi
_{\alpha}\bar{\eta}_{j}}(2,3).
\label{ward:1}
\end{equation}
Similarly, we can easily derive
\begin{equation}
G_{\Psi\eta}=G_{\hat{\Psi}\eta}=G_{\hat{\Psi}\bar{\eta}}=0.
\label{ward:2}
\end{equation}
It follows that, as in the one-loop case,
the inverse propagator is block-diagonal. Let us now
investigate whether the nonlinearity, $\delta\rho
\sum_{i}\bar{\eta}_{i}\eta_{i}$ contributes
to the self-energies in the $\Psi\hat{\Psi}$ block
($\Sigma_{\Psi\Psi},\Sigma_{\Psi\hat{\Psi}},\Sigma_{\hat{
\Psi}\Psi}$ and $\Sigma_{\hat{\Psi}\hat{\Psi}}$).
To do this, we first
note that Eqs.~(\ref{ward:1}) and (\ref{ward:2}) can
be understood as a direct consequence of the charge
conservation for $\eta ,\bar{\eta}$
imposed by the symmetry, Eq.~(\ref{sym}).
We can imagine a charge flowing through the
$\eta\bar{\eta}$ lines in the perturbative
diagrams, say from $\eta$ to $\bar{\eta}$.
The vanishing of the correlation functions with only
one external $\eta$ or $\bar{\eta}$,
Eqs.~(\ref{ward:1}) and (\ref{ward:2}), then directly
follows from the charge conservation.
The internal $\eta\bar{\eta}$
lines must then form a complete circle
following the charge flow.
The most
general diagram involving
$\eta,~\bar{\eta}$ that can contribute to
the self-energies in the $\Psi\hat{\Psi}$ block
is drawn in Fig.~2.
At one-loop order, it is just the first diagram in Fig.~1,
which vanishes \cite{psipsi}
because of the regularization, Eq.~(\ref{reg}).
We have, in fact, the same situation for the general diagram
in Fig.~2. This diagram is expressed in terms of the
integral over $l$ loop momenta and frequencies, where
$l$ is the number of the loops in the diagram.
Now we inverse Fourier transform the zeroth order
$\eta\bar{\eta}$ propagators to get the time
integral over $t_{1},\cdots ,t_{n}$,
where $n$ is the number of $G^{0}_{\eta\bar{\eta}}
(t_{i})$'s
in the diagram.
Then we can integrate over one frequency, which yields
the delta function $\delta(
\sum_{i}t_{i})$, therefore we
have the following
factor inside the integral over $l$ momenta and
$l-1$ frequencies:
\begin{equation}
\int\,\prod^{n}_{i=1}dt_{i}\:\delta (\sum^{n}_{i=1}t_{i})\,
G^{0}_{\eta\bar{\eta}}(t_{1})\cdots
G^{0}_{\eta\bar{\eta}}(t_{n})=0,
\end{equation}
which vanishes due to the
regularization, Eq.~(\ref{reg}), since
all the $G^{0}_{\eta
\bar{\eta}}(t_{i})$'s are advanced propagators.
This shows that the nonlinearity involving
the Grassmann fields does not affect
the self-energies in the $\Psi\hat{\Psi}$ block,
especially $\Sigma_{\psi\psi}=0$ even after including the
regularized Jacobian Eq.~(\ref{reg}).
We can therefore conclude that the Jacobian, included
in the DM action
as an integral over the Grassmann fields,
as in Eq.~(\ref{a3}), has no effect
on the correlation and response functions for
$\rho, {\bf g}$ and ${\bf V}$ at all orders of
perturbation theory.

\section{Formulation without {\bf V} fields}
\label{nov}

Schmitz et al. \cite{duf} have suggested that the cutoff
mechanism found by DM may be
somehow introduced artificially through the
implemetation of the constraint condition. By looking
at the original theory without the constraint
within perturbation theory, we show that this worry
is without foundation.
Starting from
the original action Eq.~(\ref{a1}), we expand $1/\rho$
as a power series in $\delta\rho$. Keeping
nonlinear terms that are
relevant to the one-loop order, we have in the action,
Eq.~(\ref{a1}),
\begin{eqnarray}
&&\sum_{j}\nabla_{j}
(\frac{g_{i}g_{j}}{\rho})\simeq
\sum_{j}\nabla_{j}(\frac{g_{i}g_{j}}{\rho_{0}})
-\sum_{j}\nabla_{j}(
\frac{g_{i}g_{j}(\delta\rho )}{\rho^{2}_{0}})+\cdots ,
\label{exp1} \\
&&\sum_{j}L_{ij}(\frac{g_{j}}{\rho})\simeq
\sum_{j}L_{ij}(\frac{g_{j}}{\rho_{0}})-
\sum_{j}L_{ij}(\frac{g_{j}(\delta\rho )}{\rho^{2}_{0}})
+\sum_{j}L_{ij}(\frac{g_{j}(\delta\rho )^{2}}{
\rho^{3}_{0}})+\cdots .\label{exp2}
\end{eqnarray}
Let us denote the results of this formulation by the
superscript II.
Only three kinds of self-energies are generated by these
nonlinearities:
$\Sigma^{(II)}_{\hat{g}_{i}\rho}({\bf q},\omega ),
\,\Sigma^{(II)}_{\hat{g}_{i}g_{j}}({\bf q},\omega )$
and $\Sigma^{(II)}_{\hat{g}_{i}\hat{g_{j}}}({\bf q},\omega )$.
Using these self-energies we can represent various
correlation and response functions. For example,
\begin{eqnarray}
&&G^{(II)}_{\rho\hat{\rho}}({\bf q},\omega )=
\frac{\omega +i\frac{\Gamma_{0}}{\rho_{0}}q^{2}-
\Sigma^{(II)}_{\hat{g}g}({\bf q},\omega )}
{D^{(II)}({\bf q},\omega )},
\label{rrh2} \\
&&G^{(II)}_{\rho\rho}({\bf q},\omega )=
\frac{q^{2}\{ 2\beta^{-1}\Gamma_{0}q^{2}-
\Sigma^{(II)}_{\hat{g}\hat{g}}({\bf q},\omega )\} }
{|D^{(II)}({\bf q},\omega )|^{2}} \label{rr2},
\end{eqnarray}
where
\begin{equation}
D^{(II)}({\bf q},\omega )=\omega^{2}-q^{2}c^{2}_{0}
-\Sigma^{(II)}_{\hat{g}\rho}({\bf q},\omega )+
i\frac{\Gamma_{0}}{\rho_{0}}q^{2}\omega-
\omega\Sigma^{(II)}_{\hat{g}g}({\bf q},\omega ).
\label{D2}
\end{equation}
We note that the number of self-energies is reduced from
seven to three compared to
the previous case, since only two variables, $\rho$
and ${\bf g}$,
are considered here. But the number of diagrams we have
to consider for each self-energy is
increased according to the appearance of
the new nonlinearities in
Eqs.~(\ref{exp1}) and (\ref{exp2}).
At one-loop order,
by comparing the nonlinear vertex structures
of the two actions, Eq.~(\ref{a1})
with the expansion
Eqs.~(\ref{exp1}), (\ref{exp2}) and
Eq.~(\ref{a2}), and by using the
relations among the linear propagators:
\begin{eqnarray}
&&G^{0}_{\rho\hat{V}}({\bf q},\omega )=
\frac{\Gamma_{0}q^{2}}{\rho_{0}}\,G^{0}_
{\rho\hat{g}}({\bf q},\omega ), \\
&&G^{0,L}_{g\hat{V}}({\bf q},\omega )=
\frac{\Gamma_{0}q^{2}}{\rho_{0}}\,
G^{0,L}_{g\hat{g}}({\bf q},\omega ), \\
&&G^{0,L}_{V\hat{V}}({\bf q},\omega )=
\frac{i}{\rho_{0}}+\frac{\Gamma_{0}q^{2}}{
\rho^{2}_{0}}\,G^{0,L}_{g\hat{g}}({\bf q},\omega ),
\end{eqnarray}
with $\Gamma_{0}$ being replaced by $\eta_{0}$ for
transverse propagators, we find
that the new diagrams for each
self-energy of (II) are
in one to one correspondence to
the diagrams of the corresponding
self-energies
of (I) that are not present in formulation (II).
For example, in Fig.~3, the first five diagrams contributing
to $\Sigma^{(II)}_{\hat{g}\rho}$ reproduce the corresponding
diagrams of $\Sigma^{(I)}_{\hat{g}\rho}$. The remaining diagrams
contain the $\hat{g}g\rho$ vertex from Eq.~(\ref{exp2}) with
$\hat{g}$ on external legs, which give us a factor
of $\frac{\Gamma_{0}q^{2}}{\rho_{0}}$ for each diagram.
Other than this factor, we find that these are the diagrams
of the self-energy $\Sigma^{(I)}_{\hat{V}\rho}$. Thus, at
one-loop order, we have
\begin{equation}
\Sigma^{(II)}_{\hat{g}\rho}({\bf q},\omega )=
\Sigma^{(I)}_{\hat{g}\rho}({\bf q},\omega )+
\frac{\Gamma_{0}q^{2}}{\rho_{0}}\,\Sigma
^{(I)}_{\hat{V}\rho}({\bf q},\omega ).
\label{rel:1}
\end{equation}
Similarly, as seen from Figs.~4 and 5, we find that
\begin{equation}
\Sigma^{(II)}_{\hat{g}g}({\bf q},\omega )=\frac{1}
{\rho_{0}}\,\Sigma^{(I)}_{\hat{g}V}({\bf q},\omega )
+\frac{\Gamma_{0}q^{2}}{\rho^{2}_{0}}\,
\Sigma^{(I)}_{\hat{V}V}({\bf q},\omega ),
\label{rel:2}
\end{equation}
\begin{eqnarray}
\Sigma^{(II)}_{\hat{g}\hat{g}}({\bf q},\omega )&=&
\Sigma^{(I)}_{\hat{g}\hat{g}}({\bf q},\omega )+
\frac{\Gamma_{0}q^{2}}{\rho_{0}}\,\Sigma^{(I)}_{\hat{g}
\hat{V}}({\bf q},\omega )
+\frac{\Gamma_{0}q^{2}}{\rho_{0}}
\,\Sigma^{(I)}_{\hat{V}\hat{g}}({\bf q},\omega )+
(\frac{\Gamma_{0}q^{2}}{\rho_{0}})^{2}\,\Sigma
^{(I)}_{\hat{V}\hat{V}}({\bf q},\omega ) \nonumber \\
&=&
\Sigma^{(I)}_{\hat{g}\hat{g}}({\bf q},\omega )
+(\frac{\Gamma_{0}q^{2}}{\rho_{0}})^{2}\,\Sigma
^{(I)}_{\hat{V}\hat{V}}({\bf q},\omega ),
\label{rel:3}
\end{eqnarray}
since $\Sigma^{(I)}_{\hat{g}
\hat{V}}({\bf q},\omega )=-\Sigma
^{(I)}_{\hat{V}\hat{g}}({\bf q},\omega )$.
Using these relations we can express the response and
correlation functions in
formulation (II), Eqs.~(\ref{rrh2}) and (\ref{rr2}),
in terms of the self-energies in
formulation (I). We first note that,
since we are dealing with one-loop corrections, we
can neglect quadratic and higher order terms in $\Sigma$'s.
Inserting Eqs.~(\ref{rel:1}) and (\ref{rel:2}) into
Eq.~(\ref{D2}), we have
\begin{equation}
D^{(II)}({\bf q},\omega )=
\rho^{-1}({\bf q},\omega )D^{(I)}({\bf q},\omega ).
\label{D12}
\end{equation}
{}From Eqs.~(\ref{rrh2}), (\ref{rr2}) and using
Eqs.~(\ref{rel:1})-(\ref{rel:3}) and (\ref{D12}), we have,
at one-loop order,
\begin{equation}
G^{(II)}_{\rho\hat{\rho}}({\bf q},\omega )=
G^{(I)}_{\rho\hat{\rho}}({\bf q},\omega ),~~~~~~~
G^{(II)}_{\rho\rho}({\bf q},\omega )=
G^{(I)}_{\rho\rho}({\bf q},\omega ),
\end{equation}
which shows that the two
formulations, (I) and (II) are {\it equivalent} to
first order. In particular,
the DM cutoff is recovered as seen from the
second term in the right hand side of Eq.~(\ref{rel:1})
and Eq.~(\ref{D12}). Therefore the cutoff is
generated not by a particular form of a
constraint condition, but by the intrinsic
$1/\rho$ nonlinearity in the problem.

The
renormalizations of various hydrodynamic
quantities for the two
formulations are also equivalent. For example,
the renormalized sound speed, for both (I) and (II),
is given by
\begin{equation}
c^{2}=c^{2}_{0}+\lim_{{\bf q},\omega\rightarrow 0}\,
\frac{1}{q}\,\Sigma_{\hat{g}\rho}({\bf q},\omega ).
\end{equation}
But, because of Eq.~(\ref{rel:1}), $\lim\frac{1}{q}
\Sigma^{(I)}_{\hat{g}\rho}=\lim\frac{1}{q}\Sigma^{
(II)}_{\hat{g}\rho}$. This quantity was evaluated
in Ref.~\cite{DM} to show that
\begin{equation}
\lim_{{\bf q},\omega\rightarrow 0}\,
\frac{1}{q}\,\Sigma^{(I)}_{\hat{g}\rho}({\bf q},\omega )
=0, \label{c:dynamic}
\end{equation}
i.e., there is no one-loop correction to the
hydrodynamic sound speed.

\section{Static Theory}
\label{sta}

We have investigated in Sec.~\ref{gras} and
Sec.~\ref{nov} the role of the Jacobian in
the dynamics governed by the Langevin equations, and
found, to lowest order, that there are no changes in the
dynamics due to the Jacobian. Let us now look at the static
equilibrium behavior.
The static equal time
averages of the fields $\psi_{i}$ are calculated
with respect to the effective Hamiltonian, Eq.~(\ref{hamil}):
\begin{equation}
\langle\psi_{i}\psi_{j}\rangle =\int\,{\cal D}
\psi\:e^{-\beta F}\psi_{i}
\psi_{j}/Z,
\end{equation}
where $Z=\int\,{\cal D}\psi\:e^{-\beta F}$.
Nonlinear corrections come from the $1/\rho$ factor in the
kinetic energy, Eq.~(\ref{kinetic}). We note that
the potential
energy part of the effective Hamiltonian $F_{u}$
appears in the Langevin equation only in the form of
$\rho\nabla^{i}(\delta F_{u}/\delta\rho )$. Therefore
a chemical potential-like term in $F_{u}$, which
is linear in $\rho$, does not affect the Langevin
equation. But, in the statics, such a term controls the
renormalizations of
the average $\langle\rho\rangle$ of the
mass density and
the sound speed $c$. To be specific,
let us take the simple example
\begin{equation}
F_{u}=\int\,d^{d}{\bf x}\:[\frac{A}{2}\rho^{2}-\mu\rho ],
\end{equation}
which gives the same Langevin equation as Eq.~(\ref{potential}).
The one-loop (i.e.\ O($\beta^{-1}$))
corrections to $\langle\rho\rangle$ can be calculated by expanding
$\rho =\langle\rho\rangle +\Delta\rho$ and integrating out
the ${\bf g}$
field. As a result, we have the following one-loop effective
action in $\Delta\rho$:
\begin{eqnarray}
F_{eff}[\Delta\rho]&=&\frac{1}{2}\{
A+\beta^{-1}\frac{d\delta_{\Lambda}(0)}{2\langle\rho
\rangle^{2}}
\}(\Delta\rho )^{2} \nonumber \\
&&+\{ A\langle\rho\rangle -\mu -\beta^{-1}
\frac{d\delta_{\Lambda}(0)}{2\langle\rho\rangle} \}
(\Delta\rho ). \label{Feef}
\end{eqnarray}
Then $\langle\rho\rangle$ is determined by setting the linear term in
$\Delta\rho$ to zero. Thus we have
\begin{eqnarray}
\langle\rho\rangle &=&\rho_{0}+\beta^{-1}\rho_{1}+\mbox{O}(\beta^{-2}), \\
\rho_{1}&=&\frac{\mu_{1}}{A}+\frac{d\delta_{\Lambda}(0)}
{2A\rho_{0}},
\end{eqnarray}
where $\mu =A\rho_{0}+\beta^{-1}\mu_{1}+\mbox{O}(\beta^{-2})$.
The sound speed is given by the
usual thermodynamic result,
\begin{equation}
c^{2}=\beta^{-1}\langle\rho\rangle\,\langle\Delta\rho\,
\Delta\rho\rangle^{-1}.
\label{c:static}
\end{equation}
The derivation of Eq.~(\ref{c:static})
from the Langevin equation follows using
the Fokker-Planck description of the problem
\cite{MRT}. The one-loop correction to the static
density-density correlation
function can easily be read off
from the quadratic term of the effective
action, Eq.~(\ref{Feef}):
\begin{equation}
\beta \langle\Delta\rho\,\Delta\rho\rangle=\frac{1}{A}-
\beta^{-1}\frac{d\delta_{\Lambda}(0)}{2A^{2}\rho^{2}_{0}}
+\mbox{O}(\beta^{-2}).
\end{equation}
Therefore,
\begin{equation}
c^{2}=c^{2}_{0}+\beta^{-1}(\mu_{1}+\frac{d\delta_{\Lambda}(0)}
{\rho_{0}})+\mbox{O}(\beta^{-2}).
\end{equation}
This is consistent with the dynamics result, Eq.~(\ref{c:dynamic}),
if we choose $\mu_{1}=-d\delta_{\Lambda}(0)/\rho_{0}$.
Thus, in some sense, the system
{\it selects} the corresponding static
limit as it evolves through
the Langevin equations.
The above discussion indicates that
any effect the Jacobian might have
on the static renormalization can be absorbed in the
chemical potential $\mu$ chosen by the Langevin equation.

\section{Conclusion}

We have shown that the Jacobian, Eq.~(\ref{jacob})
arising from the
constraint enforcing the nonlinear relation
between ${\bf g}$ and ${\bf V}$ does not
change the dynamics of Das and Mazenko.
In particular, the functional integral formulations
with and without a constraint are shown to be
equivalent at least to lowest order.
Therefore the cutoff mechanism discovered in
Ref.~\cite{DM} is a genuine effect of the
$1/\rho$ nonlinearity in the problem and not an
artifact of the particular form of the constraint
condition.

\acknowledgements
This work was supported by the National Science
Foundation Materials Research Laboratory at the
University of Chicago.

\begin{figure}
\caption{The one-loop diagrams generated by the nonlinearity
due to the Jacobian. The Jacobian is regularized such that the
first diagram vanishes.}
\end{figure}
\begin{figure}
\caption{The most general diagram
generated by the nonlinearity
due to the Jacobian
without external Grassmann fields.
The arrow indicates the charge
flow.}
\end{figure}
\begin{figure}
\caption{The detailed correspondence between the one-loop
diagrams contributing to $\Sigma^{(II)}_{\hat{g}\rho}$
and $\Sigma^{(I)}_{\hat{g}\rho}+\frac{\Gamma_{0}
q^{2}}{\rho_{0}}\Sigma^{(I)}_{\hat{V}\rho}$. See
Eq.~(\protect\ref{rel:1}).}
\end{figure}
\begin{figure}
\caption{The detailed correspondence between the one-loop
diagrams contributing to the self-energies in Eq.~(\protect\ref{rel:2}).}
\end{figure}
\begin{figure}
\caption{The detailed correspondence between the one-loop
diagrams contributing to the self-energies in Eq.~(\protect\ref{rel:3}).}
\end{figure}

\end{document}